\documentstyle[prl,aps,twocolumn,floats,amssymb,epsfig]{revtex}

\def\SU{{\rm{SU}}}

\newcommand{\mb}[1]{\ifmmode#1\else\mbox{$#1$}\fi}

\newcommand\de{\mb{\delta}}

\newcommand\la{\mb{\lambda}}

\newcommand\om{\mb{\omega}}

\newcommand\Om{\mb{\Omega}}

\newcommand\calD{\mb{{\cal D}}}

\newcommand\calH{\mb{{\cal H}}}
\newcommand\calI{\mb{{\cal I}}}

\newcommand\calL{\mb{{\cal L}}}

\newcommand\calV{\mb{{\cal V}}}

\newcommand{\beq}{\begin{equation}}
\newcommand{\eeq}{\end{equation}}
\newcommand{\nn}{\nonumber}
\newcommand{\bea}{\begin{eqnarray}}
\newcommand{\eea}{\end{eqnarray}}

\newcommand{\pderiv}[2]{\frac{\partial {#1}}{\partial {#2}}}

\newcommand{\rhat}{\mb{\hat{\bm r}}}

\newcommand{\gsim}
{\raise.3ex\hbox{$\;>$\kern-.75em\lower1ex\hbox{$\sim$}$\:$}}
\newcommand{\lsim}
{\raise.3ex\hbox{$\;<$\kern-.75em\lower1ex\hbox{$\sim$}$\:$}}
\newcommand{\ts}{\textstyle}

\newcommand{\half}{{\ts \frac{1}{2}}}
\newcommand{\quarter}{{\ts \frac{1}{4}}}

\newcommand{\bm}[1]{{\mbox{\boldmath $#1$}}}

\renewcommand{\O}{{\rm O}}

\begin{document}
\renewcommand{\thesubsection}{\arabic{subsection}} 
\draft


\twocolumn[\hsize\textwidth\columnwidth\hsize\csname @twocolumnfalse\endcsname
\author{Nathan\  F.\  Lepora\footnotemark}
\title{Some problems when calculating the 
quantum corrections\\ to the classical 't Hooft-Polyakov monopole} 
\date{Submitted 24 March 2002; accepted April 27, 2002}
\maketitle

\begin{abstract}
We examine whether the analysis of quantum corrections for the kink soliton
carries over to the 't~Hooft-Polyakov monopole. For the kink, it is central
that the quantum fluctuations are eigenmodes of a Hermitian operator. 
For the monopole, we show the analogous operator is not Hermitian. This 
property raises
some questions about the quantization procedure for a 't~Hooft-Polyakov
monopole.
\end{abstract}
\pacs{Published in: Phys. Lett. {\bf B}524 (2002) 338.}]


\subsection{Introduction}

While quantum corrections to the classical 't Hooft-Polyakov 
monopole~\cite{hooft} have not been explicitly found, they have
been for another soliton --- the kink~\cite{kink}. It seems
reasonable that the calculations for the monopole and kink should be similar. 
Assuming such a similarity, a
suitable framework would start with the classical monopole,
then determine the quantum fluctuations and, from these, derive
the quantum corrections. 

In this paper we draw attention to a feature of the quantum 
fluctuations that significantly differs between the monopole and
kink. While closer examination may reveal this 
difference to be merely a technical issue, we think it could
reflect a subtlety in the way semi-classical solitons behave.

To introduce our arguments, we briefly describe the quantum fluctuations
around a kink soliton. Then we cover the analogous situation for
the 't~Hooft-Polyakov monopole and indicate some problems. 
Finally, we discuss the cause of these problems.

\subsection{The kink and its quantum corrections}

The kink soliton~\cite{kink} is a classical solution to a (1+1)-dimensional 
scalar field theory with Lagrangian density
\beq
\calL[\Phi]=\half\dot\Phi^2-\half\Phi'^2-V[\Phi],
\hspace{.9em}
V[\Phi]=\quarter\la\left(\Phi^2-v^2\right)^2\!\!\!,
\eeq
and has a classical scalar field profile
\beq
\label{kink}
\Phi_{\rm cl}(x)=\frac{m}{\sqrt{\la}}\tanh\frac{mx}{\sqrt 2},
\hspace{1em}m^2=\la v^2.
\eeq
This kink has a classical mass defined from the Hamiltonian
\beq
\label{Mcl}
M_{\rm cl}=H[\Phi_{\rm cl}]=\int\!d^3x\,\calH[\Phi_{\rm cl}]
=\frac{2\sqrt{3}m^3}{3\la}.
\eeq
Here the Hamiltonian density is defined in the usual way
\beq
\label{Hper}
\calH[\Phi,\Pi]=\Pi\dot\Phi-\calL
=\half\Pi^2+\half\Phi'^2+V[\Phi],
\eeq
where $\Pi=\partial\calL/\partial\dot\Phi=\dot\Phi$ is the canonical
momentum.

Quantum corrections to the classical kink originate
from fluctuations $\phi$ around the classical background
\beq
\label{kper}
\Phi=\Phi_{\rm cl}+\phi.
\eeq
Then the Hamiltonian for the fluctuations is found by 
substituting (\ref{kper}) into (\ref{Hper}) and expanding to $\O(\phi^3)$
\bea
\label{kh}
H[\phi,\pi]=M_{\rm cl} + \int d^3r \left[\half\,\pi^2+
\half\,\phi\left(-\partial_x^2 + \calV\right)\phi\right]
\eea
with $\pi=\dot\phi$ the canonical momentum and $\calV$ a potential for the
fluctuations
\beq
\label{kv}
\calV(x)=3m^2\tanh^2\frac{mx}{\sqrt 2}-m^2.
\eeq
This potential is a well with asymptote $\calV(\infty)=2m^2$.

It is crucial that the following operator is Hermitian:
\beq
\label{kherm}
\int\! d^3x\,\,\psi\!\left(-\partial_x^2 + \calV\right)\!\phi =
\int\! d^3x\,\,\phi\!\left(-\partial_x^2 + \calV\right)\!\psi.
\eeq
This relation follows from partial integration, neglecting surface terms.
There is, therefore, a complete set of orthonormal eigenmodes
\beq
\label{sch}
\left(-\partial_x^2 + \calV\right)\phi_s=\om_s^2\phi_s,
\hspace{1.5em}
\int\! d^3x\,\phi_s\phi_t=\de_{st}.
\eeq
Expanding the fluctuation field as a sum $\phi=\ts\sum_s q_s\phi_s$ 
over these eigenmodes, we find
\beq
H(p_s,q_s) = M_{\rm cl} + \ts\sum_s\left[p_s^2+\om_s^2q_s^2\right]+\cdots,
\eeq
where $q_s= \int d^3x\,\phi\phi_s$; $p_s=\dot q_s$; the classical mass
$M_{\rm cl}$ is from
(\ref{Mcl}) and the dots represent $\O(\phi^3)$ terms.

Therefore the Hamiltonian (\ref{kh}) becomes a sum
over individual Hamiltonians $H(p_s,q_s)$ that describe
Harmonic oscillators. These oscillators can be
quantized with Heisenberg's commutation relation
$[p_s,q_t]=i\hbar\de_{st}$, giving
\beq
\label{Hn}
H(n_{\om_s})=M_{\rm cl} + \ts\sum_s(n_{\om_s}+\half)\hbar\om_s + \cdots.
\eeq
Here $n_{\om_s}$ is the occupation number of the Harmonic oscillator 
with frequency $\om_s$.

In conclusion, the fluctuations around the classical 
kink can be written as a sum over eigenmodes that are
stationary states of a time-independent Schr\"odinger equation (\ref{sch}).
When these states are quantized, the meson spectrum
has energies $(n_{\om_s}+\half)\hbar\om_s$. Then
the quantum correction to the classical kink mass is a sum
over the 
differences between the meson's zero-point and vacuum energies
\beq
M = M_{\rm cl} + \ts\sum_s \half\hbar\,\de\om_s + \cdots.
\eeq
Regularization techniques are used to evaluate this sum, which
gives a finite value of order $\hbar m$.

In the above discussion, there are a couple of points that will be 
relevant to our later arguments:\\
(a) The operator in (\ref{kherm}) is Hermitian --- then
the fluctuations can be expanded over a complete set of 
eigenmodes.\\
(b) The quantum fluctuations solve a time-independent 
Schr\"odinger equation (\ref{sch}) --- 
giving a connection between quantum field theory and quantum mechanics.

Later on in this paper, we compare these points with the behaviour
of the 't~Hooft-Polyakov monopole.

\subsection{The 't~Hooft-Polyakov monopole}

Now we attempt to repeat the above arguments for a 't~Hooft-Polyakov monopole.
Our aim is to see whether there are any significant differences
between the kink and monopole.

The 't~Hooft-Polyakov monopole~\cite{hooft} is a classical solution to 
a (3+1)-dimensional $\SU(2)$ scalar-gauge field theory with Lagrangian density
\bea
\label{lag}
&&\calL[\Phi_a,A^i_a] = 
-\quarter F^{\mu\nu}_aF_{a\mu\nu} + \half D^\mu\Phi_a D_\mu\Phi_a
- V[\Phi_a],\nn\\
&&\hspace{4em}F^{\mu\nu}_a=\partial^\mu A^\nu_a-
\partial^\nu A^\mu_a+e\,\epsilon_{abc}A^\mu_bA^\nu_c,\nn\\
&&\hspace{4em}
D^\mu\Phi_a=\partial^\mu\Phi_a + e\,\epsilon_{abc}A^\mu_b\Phi_c,\\
&&\hspace{4em}
V[\Phi_a]=\quarter\la(\Phi_a\Phi_a-v^2)^2,\nn
\eea
and has classical scalar and gauge fields
\beq
\label{jz}
(\Phi_a)_{\rm cl} = \frac{H(r)}{er}\hat r_a,\hspace{1em}
(A^i_a)_{\rm cl} = \frac{K(r)-1}{er}\,\epsilon_{iab}\hat r_b
\eeq
in the radial $\epsilon_{abc}\Phi_b\rhat_c$ and temporal $A^0_a=0$ gauge. 
Single-valuedness requires $H(0)=K(0)-1=0$, 
while the asymptotic boundary conditions are $H/r\rightarrow ev$ and 
$K\rightarrow 0$ as $r\rightarrow\infty$.

This monopole has magnetic charge $4\pi/e$ and vanishing
electric charge, both of which are defined in a gauge-invariant way
\bea
\label{g}
g=\frac{1}{v}\int_{S_\infty}\!\!\!\! dS^i\,\Phi_a B^i_a= \frac{4\pi}{e},\,\,\,
q=\frac{1}{v}\int_{S_\infty}\!\!\!\! dS^i\,\Phi_a E^i_a= 0.\!\!
\eea
These charges are defined by the non-Abelian electric and magnetic fields,
$E^i_a=F^{0i}_a$ and $B^i_a=-\half\epsilon_{ijk}F^{jk}_a$. 

The classical mass of the monopole is defined from the Hamiltonian~\cite{note1}
\bea
\label{mcl}
M_{\rm cl}=H\!\left[(\Phi_a)_{\rm cl},(A^i_a)_{\rm cl}\right]
=\int\!d^3x\,\calH\!\left[(\Phi_a)_{\rm cl},(A^i_a)_{\rm cl}\right]\!,
\eea
which is specified by the Hamiltonian density
\beq
\label{H}
\calH[\Phi_a,\Pi_a;A^i_a,\Pi^i_a]=\Pi_a\dot\Phi_a +
\Pi^i_a\dot A^i_a -\calL[\Phi_a,A^i_a]
\eeq
with canonical momenta
$\Pi_a=\partial\calL/\partial\dot\Phi_a=\dot\Phi_a$ for the scalar field
and $\Pi^i_a=\partial\calL/\partial\dot A^i_a=E^i_a$ for the gauge field.

\subsection{Fluctuations around the monopole}

Quantum corrections to the classical monopole mass 
originate from the fluctuations $\phi_a$ and $a^i_a$ around the classical
fields $(\Phi_a)_{\rm cl}$ and $(A^i_a)_{\rm cl}$.
To ease the following calculations, we consider these fluctuations 
in the radial $\epsilon_{abc}\Phi_b\hat r_c=0$ and temporal $A^0_a=0$ gauge
\bea
\label{fluc}
\Phi_a = (\Phi_a)_{\rm cl} + \phi\,\hat r_a,\hspace{1em}
A^i_a = (A^i_a)_{\rm cl} + a^i_a.
\eea
This gauge choice reduces the scalar field to only one degree-of-freedom.

To describe these fluctuations, we construct the Hamiltonian from the 
fluctuation fields and their canonical momenta
[like with the kink in eqs.~(\ref{kper}-\ref{kv})].
These canonical momenta are related to the fluctuations 
(\ref{fluc}) by
\bea
\Pi_a=(\Pi_a)_{\rm cl} + \pi\hat r_a=\pi\hat r_a,\hspace{.5em}
\Pi^i_a=(\Pi^i_a)_{\rm cl} + \pi^i_a=\pi^i_a,\!
\eea
where $\pi=\dot\phi$ and $\pi^i_a=-\dot a^i_a$ in the temporal gauge.
Then the Hamiltonian separates into
zeroth-, linear- and quadratic-order terms in the fluctuations
$\phi$, $\pi$, $a^i_a$, $\pi^i_a$:
\bea
\label{Hdec}
H[\Phi_a,\Pi_a;A^i_a,\Pi^i_a]=H_{\rm cl} + H_{\rm lin} + H_{\rm qu}+\cdots.
\eea
Clearly,
\beq
H_{\rm cl}=H\!\left[(\Phi_a)_{\rm cl},(A^i_a)_{\rm cl}\right]=M_{\rm cl},
\eeq
while it can be shown that
\beq
H_{\rm lin}=0.
\eeq
This linear Hamiltonian vanishes by $\calL_{\rm lin}=0$, which is
because
the monopole (\ref{jz}) solves the equations of motion and is then
a stationary point of the action.

Therefore we are left with only a quadratic-order Hamiltonian
$H_{\rm qu}$ for the fluctuations, which has a density
\beq
\label{Hd}
\calH_{\rm qu}=\pi_a\pi_a+\pi^i_a\pi^i_a
-\calL_{\rm qu}.
\eeq
Here $\calL_{\rm qu}$ contains the quadratic-order  fluctuation terms in the 
Lagrangian density
\beq
\calL_{\rm qu} = 
- \quarter \left(F^{\mu\nu}_aF_{a\mu\nu}\right)_{\rm qu} 
+ \half \left(D^\mu\Phi_a D_\mu\Phi_a\right)_{\rm qu}
- V_{\rm qu}.
\eeq
We now determine each of these terms.

First, we consider the components of 
$-\quarter F^{\mu\nu}_aF_{a\mu\nu}$ that are quadratic in the 
fluctuations. These components are more easily found from
the non-Abelian electric and magnetic fields
\bea
-\quarter(F^{\mu\nu}_aF_{a\mu\nu})_{\rm qu}&=&
\half(E^i_a)_{\rm lin}(E^i_a)_{\rm lin} + 
(E^i_a)_{\rm cl}(E^i_a)_{\rm qu} \nn\\
\label{ff}
&&\hspace{-1em}-\,\half(B^i_a)_{\rm lin}(B^i_a)_{\rm lin} -
(B^i_a)_{\rm cl}(B^i_a)_{\rm qu}.
\eea
Here the zeroth-, linear- and quadratic-order fields are found
from $E^i_a=F^{0i}_a$ and 
$B^i_a=-\half\,\epsilon_{ijk}F^{jk}_a$. For the non-Abelian electric field,
we find
\beq
\label{e1}
(E^i_a)_{\rm lin}=-\dot a^i_a=\pi^i_a,\hspace{1em}
(E^i_a)_{\rm qu}=0;
\eeq
while for the non-Abelian magnetic field,
\bea
\label{b1}
(B^i_a)_{\rm lin}&=&\epsilon_{ijk}\partial^ja^k_a
+e\,\epsilon_{ijk}\epsilon_{abc}(A^j_b)_{\rm cl}\,a^k_c,\\
\label{b2}
(B^i_a)_{\rm qu}&=&\half e\,\epsilon_{abc}\epsilon_{ijk}a^j_ba^k_c.
\eea
To keep things simple,  we shall write the classical fields as
$(A^i_a)_{\rm cl}$ and $(B^i_a)_{\rm cl}$.
Substituting (\ref{e1}-\ref{b2}) into (\ref{ff}), we then find
\bea
&&-\quarter(F^{\mu\nu}_aF_{a\mu\nu})_{\rm qu}= \pi^i_a\pi^i_a
-\half(B^i_a)_{\rm lin}(B^i_a)_{\rm lin} \nn\\
&&\hspace{13em}- (B^i_a)_{\rm cl}(B^i_a)_{\rm qu},
\eea
where upon partial integration and then neglecting the surface terms
\bea
\label{blinsq}
(B^i_a)_{\rm lin}(B^i_a)_{\rm lin}&=&
a^i_a\left(\nabla^2\de^{ij}-\partial^j\partial^i\right)a^j_a \nn\\
&+& a^i_a\left[
e^2\epsilon_{ikm}\epsilon_{jlm}\epsilon_{acd}\epsilon_{bce}
(A^k_d)_{\rm cl}(A^l_e)_{\rm cl}\right]a^j_b \nn\\
&-& a^i_a\left[
2e\,\epsilon_{ikm}\epsilon_{abc}\epsilon_{jlm} 
(A^k_c)_{\rm cl}\partial^l\right]a^j_b,\\
(B^i_a)_{\rm cl}(B^i_a)_{\rm qu}&=&
\half e\,\epsilon_{abc}\epsilon_{ijk}(B^k_c)_{\rm cl}a^i_aa^j_b.
\eea
Note the linear-derivative term in (\ref{blinsq}) --- this will be
troublesome later.

Next, we consider the components of $\half D^\mu\Phi D_\mu\Phi$
that are quadratic in the fluctuations. These components are found by
expanding
\bea
\label{dd}
\half(D^\mu\Phi_aD_\mu\Phi_a)_{\rm qu}
&=&\half\pi_a\pi_a - \half(D^i\Phi_a)_{\rm lin}(D^i\Phi_a)_{\rm lin}\nn\\
&&\hspace{2em}-\,(D^i\Phi_a)_{\rm cl}(D^i\Phi_a)_{\rm quad}.
\eea
Here we write $\phi_a=\phi\,\hat r_a$, which gives $D^0\phi_a=\dot\phi_a=\pi_a$
in a temporal gauge. Similarly, the linear- and
quadratic-order components of the covariant derivative are
\bea
(D^i\Phi_a)_{\rm lin}&=&\partial^i\phi_a
+e\,\epsilon_{abc}a^i_b(\Phi_c)_{\rm cl}
+e\,\epsilon_{abc}(A^i_b)_{\rm cl}\phi_c,\\
(D^i\Phi_a)_{\rm qu}&=&e\,\epsilon_{abc}a^i_b\phi_c.
\eea 
To keep things simple, we shall write the classical component
of the covariant derivative as $(D^i\Phi_a)_{\rm cl}$. Then the
expressions in (\ref{dd}) are
\bea
&&(D^i\Phi_a)_{\rm lin}(D^i\Phi_a)_{\rm lin}
=-\phi_a\nabla^2\phi_a\nn\\
&&\hspace{3em}
+e^2\epsilon_{abc}\epsilon_{ade}\left[(A^i_b)_{\rm cl}(A^i_d)_{\rm cl}
\phi_c\phi_e + (\Phi_c)_{\rm cl}(\Phi_e)_{\rm cl} a^i_ba^i_d\right]\nn\\
&&\hspace{3em}
+2e\,\partial^i\phi_a\epsilon_{abc}\left[a^i_b(\Phi_c)_{\rm cl}+
(A^i_b)_{\rm cl}\phi_c\right]\nn\\
&&\hspace{3em}
+2e^2\epsilon_{abc}\epsilon_{ade}(\Phi_c)_{\rm cl}(A^i_d)_{\rm cl}
a^i_b\phi_e,\\
&&(D^i\Phi_a)_{\rm cl}(D^i\Phi_a)_{\rm qu}
= e\,\epsilon_{abc} (D^i\Phi_a)_{\rm cl}\,a^i_b\phi_c.
\eea
Again we find several terms that are linear in the field derivatives. 
However, this time the radial gauge
expression $\partial^i\phi_a=r\partial^i(\phi/r)\hat r_a + (\phi/r)\de_{ia}$ 
implies
\bea
e\,\epsilon_{abc}(A^i_b)_{\rm cl}(\partial^i\phi_a)\phi_c
&=&e\,\epsilon_{ibc}(A^i_b)_{\rm cl}\,\hat r_c\phi^2/r,\\
e\,\epsilon_{abc}(\Phi_c)_{\rm cl}(\partial^i\phi_a)a^i_b
&=&e\,\epsilon_{ibc}(\Phi_c)_{\rm cl}\,a^i_b\phi/r.
\eea
Therefore these terms are not problematic.

Finally, there are also quadratic fluctuation terms in the scalar potential,
\beq
\label{vqu}
V_{\rm qu}[\phi]=\ts\frac{3}{2}\la(\Phi_a)_{\rm cl}(\Phi_a)_{\rm cl}\phi^2
-\half\la v^2\phi^2.
\eeq
These are found by substituting (\ref{jz}) into $V[\Phi]$.

Hence the final expression for the Hamiltonian is found by 
putting everything between (\ref{mcl}) and (\ref{vqu}) together into
\beq
\label{hm}
H=M_{\rm cl}+H_{\rm qu}[\phi,\pi;a^i_a,\pi^i_a]+\cdots.
\eeq
This $H_{\rm qu}$ splits into scalar, gauge and mixed parts
\beq
H_{\rm qu}=H_{\rm qu}[\phi,\pi] + H_{\rm qu}[a^i_a,\pi^i_a]
+I_{\rm qu}[\phi,\pi;a^i_a,\pi^i_a],
\eeq
specified by a Hamiltonian density for scalar fluctuations
\bea
\label{hphi}
\calH_{\rm qu}[\phi,\pi]&=&\half\pi_a\pi_a - \half \phi_a\nabla^2\phi_a
+\half\calV_\phi\,\phi^2,\\
\calV_\phi(r)&=&
e^2\epsilon_{abc}\epsilon_{ade}(A^i_b)_{\rm cl}(A^i_d)_{\rm cl}
+3\la(\Phi_a)_{\rm cl}(\Phi_a)_{\rm cl}\nn\\
&&+\,2e\,\epsilon_{ibc}(A^i_b)_{\rm cl}\hat r_c/r-\la v^2 ;
\eea
a Hamiltonian density for gauge fluctuations
\bea
\label{ha}
\calH_{\rm qu}[a^i_a,\pi^i_a]&=&\half\pi^i_a\pi^i_a -
\half a^i_a\left(\nabla^2\de^{ij}-\partial^j\partial^i\right)a^j_a\nn\\
&&\hspace{-2em}+\,\half a^i_a\calV^{ij}_{ab}a^j_b 
+\half a^i_a\calD^{ij}_{ab}a^j_b,\\
\calV^{ij}_{ab}(r)&=&
e^2\epsilon_{ikm}\epsilon_{jlm}\epsilon_{bcd}\epsilon_{ace}
(A^k_d)_{\rm cl}(A^l_e)_{\rm cl}\\
&&\hspace{-2em}+\,e\,\epsilon_{abc}\epsilon_{ijk}(B^k_c)_{\rm cl}
+e^2\,\epsilon_{bcd}\epsilon_{aed} 
(\Phi_c)_{\rm cl}(\Phi_e)_{\rm cl}\de^{ij},\nn\\
\label{dij}
\calD^{ij}_{ab}&=&
-2e\,\epsilon_{ikm}\epsilon_{abc}\epsilon_{jlm}
(A^k_c)_{\rm cl}\partial^l;
\eea
and a scalar-gauge mixing term
\bea
\label{hphia}
\calI_{\rm qu}[\phi,\pi;a^i_a,\pi^i_a]&=&\calV^i_aa^i_a\phi,\\
\calV^i_a
&=&e^2\epsilon_{abc}\epsilon_{bde}(\Phi_c)_{\rm cl}(A^i_d)_{\rm cl}
+e\,\epsilon_{abc}(D^i\Phi_b)_{\rm cl} \nn\\
\label{hlast}&&+\, e\,\epsilon_{aic}(\Phi_c)_{\rm cl}/r. 
\eea
Collectively, the above equations (\ref{hm}-\ref{hlast})
are analogous to (\ref{kh}) and (\ref{kv}) for the kink.

\subsection{Problems with quantization}

We now comment on these equations and indicate some differences from the
kink in section 2.

The Hamiltonian (\ref{hm}) has a similar layout to that
for the kink: with one part the classical mass and the other a quadratic
Hamiltonian for the fluctuations. Therefore one might expect the 
next step would be to expand 
\beq
\label{modes}
\phi=\sum_s\phi_sq_s,\hspace{1em}
a^i_a=\sum_s(a^i_a)_sQ_s
\eeq
with $\phi_s$, $(a^i_a)_s$ the eigenfunctions of
\bea
\left(-\nabla^2 + \calV_\phi\right)\phi_s&=&\om_s^2\phi_s,\\
\left(\de_{ab}\partial^j\partial^i-\nabla^2\de_{ab}\de^{ij}
+\calV^{ij}_{ab} +\calD^{ij}_{ab}\right)(a^i_a)_s&=&\Om_s^2(a^j_b)_s.
\eea
Afterwards we could write each quadratic-order
Hamiltonian as a sum over Harmonic oscillators. These oscillators could
then be quantized.

However, there are a couple of obstacles to this procedure.

First, there is a a scalar-gauge mixing term (\ref{hphia}). Then
the eigenmodes are not separate scalar and gauge modes,
but are instead a mixture of the two. 
This problem is mainly technical and makes any calculation more difficult.

Secondly, the linear derivative term $\calD^{ij}_{ab}$ in (\ref{dij}) means
the operator
\beq
\label{op}
-\nabla^2\de_{ab}\de^{ij}+\de_{ab}\partial^j\partial^i
+\calV^{ij}_{ab}+\calD^{ij}_{ab}
\eeq
is not Hermitian because
\beq
\label{notherm}
\int\! d^3x\,a^i_a\calD^{ij}_{ab} b^j_b \neq 
\int\! d^3x\, b^i_a\,\calD^{ij}_{ab} a^j_b.
\eeq
In fact, examination of (\ref{notherm}) reveals Hermiticity only holds
for gauge perturbations that satisfy
\beq
(A^i_a)_{\rm cl}\partial^la^j_b=
(\partial^iA^l_a)_{\rm cl}a^j_b
+(A^l_a)_{\rm cl}\partial^ia^j_b.
\eeq
One cannot expect such an arbitrary constraint to hold for general gauge
perturbations.

Hence it seems there is a fundamental problem with quantizing 
the fluctuations. For quantization, the fluctuation fields should
be expanded over the eigenmodes. Then the Hamiltonian can be expanded as
a sum over Harmonic oscillators $H=\sum_sp_s^2+\om_s^2q_s^2$, which can then 
be quantized with $[p_s,q_t]=i\hbar\de_{st}$. But if the operator
(\ref{op}) is not Hermitian there are problems with the eigenmode expansion.
In particular:\\
\noindent (a) The eigenmodes may no longer be complete --- then the 
expansions in (\ref{modes}) cannot be justified since there may be 
gauge fluctuations without an eigenmode sum.\\
\noindent (b) The eigenmodes may no longer be orthogonal --- then
the coefficients $Q_s$ may not be written as $Q_s=\int\! d^3x\, aa_s$.
This problem is less serious than (a).\\
\noindent (c) We lose the connection between quantum field theory and
quantum mechanics.

Without fully examining the eigenmodes of (\ref{op}), which is
a very difficult problem, one cannot really say how the analysis 
should be modified. It is possible that if there are fewer quantum 
fluctuations then there could be a smaller quantum mass-correction. 
In addition, there could be other effects that are difficult
to anticipate.

We should mention that a more detailed 
calculation for the 't~Hooft-Polyakov monopole
will be presented elsewhere~\cite{mequant}. In that paper, many 
expressions given here will be presented in terms of the 
profile functions, rather than the classical fields. We will also
estimate the quantum correction to the monopole mass.

\subsection{Discussion}

To complete our discussion, we examine when this loss of Hermiticity can occur
in other situations by taking
two illustrative examples: purely scalar field theory and Maxwell theory.

For the first example, we consider a scalar field theory with potential
$V[\Phi]$
\beq
\calL[\Phi]=\half\partial_\mu\Phi\partial^\mu\Phi-V[\Phi].
\eeq
We take an arbitrary classical background $\Phi_{\rm cl}$ and introduce
a fluctuation field $\phi$ with
\beq
\Phi=\Phi_{\rm cl}+\phi.
\eeq
While $\phi\ll\Phi_{\rm cl}$, the potential is to $\O(\phi^3)$
\bea
V[\Phi]=V[\Phi_{\rm cl}]+\pderiv{V}{\Phi}[\Phi_{\rm cl}]\,\phi
+\frac{1}{2!}\frac{\partial^2V}{\partial\Phi^2}[\Phi_{\rm cl}]\,\phi^2.
\eea
Hence the quadratic terms are
\beq
\calL_{\rm qu}[\phi]=\half\partial_\mu\phi\partial^\mu\phi
-\frac{1}{2!}\frac{\partial^2V}{\partial\Phi^2}[\Phi_{\rm cl}]\,\phi^2,
\eeq
and the quadratic Hamiltonian is
\beq
\calH_{\rm qu}[\phi,\pi]=
\half\pi^2 + \half \phi \left(\!-\nabla^2+
\frac{\partial^2V}{\partial\Phi^2}[\Phi_{\rm cl}]\right)\phi.
\eeq
In this purely scalar field theory, the relevant, bracketed, operator
is Hermitian. Therefore the loss of Hermiticity for the monopole
appears to be a gauge sector effect.

In our second example, we consider a similar analysis to above, but
for Maxwell theory
\beq
\label{max}
\calL=-\quarter F^{\mu\nu}F_{\mu\nu},\hspace{1em}
F^{\mu\nu}=\partial^\mu A^\nu-\partial^\nu A^\mu.
\eeq
Again we take an arbitrary classical background $A^\mu_{\rm cl}$ and 
introduce a fluctuation field $a^\mu$ with
\beq
A^\mu=A^\mu_{\rm cl}+a^\mu.
\eeq
Substituting this into (\ref{max}), we find a quadratic Lagrangian density
\beq
\calL_{\rm qu}[a^\mu]=-\quarter f^{\mu\nu}f_{\mu\nu},\hspace{1em}
f^{\mu\nu}=\partial^\mu a^\nu - \partial^\nu a^\mu.
\eeq
Because of the linearity of Maxwell theory, the fluctuation field decouples
from the classical background. In Maxwell theory
there is no problem with quantizing the fluctuations.

Therefore the loss of Hermiticity in the fluctuation
Hamiltonian appears to be caused by the non-Abelian nature of the gauge theory.
In purely scalar or Abelian theories no such loss of Hermiticity
occurs.

For a final comment, we note this effect could be relevant to other classical
solutions that contain non-Abelian gauge fields --- 
an example being Skyrmions in the gauged sigma model.

\renewcommand\footnoterule\ 
\footnotetext[1]{\ email: n$\_$lepora@hotmail.com}



\begin{thebibliography}{99}

\bibitem{hooft}
G.~'t Hooft,
Nucl.\ Phys.\  {\bf B79} (1974) 276;
A.~M.~Polyakov, 
JETP Lett. 20 (1974) 194.

\bibitem{kink}
R.~Rajamaran, {\em Solitons and Instantons}, North-Holland (1996); see
also T.~Weidig, hep-th/9912005.

\bibitem{note1}
In closed form $M_{\rm cl}= 4\pi/e^2 \int dr [(K')^2
+\half(H'-H/r)^2+ K^2H^2/r^2+ \half(K^2-1)^2/r^2
+ \quarter\la(H^2-e^2v^2r^2)^2/r^2]$.

\bibitem{mequant}
N.~Lepora,
submitted to Nucl. Phys. {\bf B}.

\end{thebibliography}
\end{document}